\documentclass[aps,prl,groupedaddress,showpacs,twocolumn,amsmath,amssymb]{revtex4}

\def\beginwide{
        \end{multicols} \vspace*{-0.5cm} \noindent
        \rule{3.5in}{.1mm}\rule{.1mm}{5mm} \widetext \medskip }

\def\endwide{
        \hspace*{3.35in}~\rule[-5mm]{.1mm}{5mm}\rule{3.5in}{.1mm}
        \begin{multicols}{2} \vspace*{-1.0cm} \noindent }

\usepackage{graphicx}
\usepackage{dcolumn}

\begin{document}

\title{Criticality of a dissipative self-organized branching process in a dynamic population\\ 
}
\author{Dranreb Earl Juanico, Christopher Monterola, and Caesar Saloma}
\homepage[Fax:]{+632 9280296}
\email{djuanico@nip.upd.edu.ph}
\affiliation{National Institute of Physics, University of the Philippines\\
Diliman, Quezon City, Philippines 1101}

\date{\today}

\begin{abstract}
We derive a general formulation of the self-organized branching process by
considering sandpile dynamics in an evolving population characterized by ``birth"
(excitation) and ``death" (de-excitation)
of active sites ($z=1$). New active sites are born in empty sites ($z=0$) with a probability of $\eta$,
whereas active sites die, thus becoming empty, with a probability $\lambda$. 
Subsequently, when an active site becomes unstable ($z=2$), it topples by transferring
two grains to two randomly chosen sites with probability $\alpha$ or, by transferring only one
grain to a randomly selected site (while retaining the other) with probability
$\beta=1+\frac{\lambda}{\eta}-2\alpha$, thus remaining active after toppling. We show that when 
sandpile dynamics occurs in an evolving population, self-organized criticality, characterized by
a power-law avalanche size distribution with exponent $\tau_s=3/2$ and power-law avalanche 
duration distribution with exponent $\tau_T=2$ at very high dimension $n >> 1$, is
achieved even in the presence of dissipation ($\epsilon = 1-\alpha - \beta > 0$), contrary to 
previous claims.
\end{abstract}

\pacs{89.75.Da, 89.75.Hc, 89.65.Cd}

\maketitle
The study of self-organized criticality (SOC) has received widespread attention because
of its inherent simplicity of simulating scaling phenomena in a multitude of real
systems. The sandpile model of Bak et al. is an implementation of SOC. It had a 
significant impact to the research community because of its capability of simulating
power-law distributions observed in a lot of systems in nature. How exactly the system approaches
the SOC state is not so clear. A mean-field approach sheds light on this issue by providing a
qualitative description of the spontaneous emergence of criticality in the system. Recently,
Lauritsen, Zapperi and Stanley~\cite{lzs95} proposed an analytic method that treats the
sandpile dynamics as a branching process -- a well-founded mathematical theory that was 
initially applied to understanding spread of epidemics~\cite{harris63}. The model was
aptly named \emph{self-organized branching process} or SOBP, which is a very useful
way of analyzing the sandpile dynamics. 

The SOBP relaxation process is conservative such that the transferred ``energy" (grain) is equal
to the ``energy" lost by a relaxing site. Consequently, the system self-organizes to a critical steady state 
that is characterized by a 
power-law distribution for avalanche sizes (with exponent $\tau_s = 3/2$) and for  
avalanche lifetime distribution (with exponent $\tau_T = 2$).
The same group later demonstrated that any degree of dissipation in the relaxation rules leads to a
subcritical steady state that manifests with a characteristic size in the avalanche distribution
(power law followed by a significant drop-off after some characteristic avalanche size)~\cite{lzs96}.
A probability $\epsilon$ that an unstable active site relaxes by dissipating grains out of
the system was defined. Only at $\epsilon = 0$ is the self-organized branching process become
critical. The results apparently suggest that SOC can only occur in a conservative wherein the
relaxation process is non-dissipative.
Therefore, the existence of SOC in nonconservative models
would be highly desirable in this context, since, in practice, some type of dissipation is always 
present in nature~\cite{cp00}. An actual example of a dissipative system displaying SOC is an
earthquake, which is successfully modelled by the Olami-Feder-Christensen (OFC) model. The OFC
model sparked a wide interest because the relaxation process is dissipative. However, the 
criticality as observed in the power-law earthquake intensity distribution was recently questioned
by de Carvalho and Prado~\cite{cp00}, who strongly claimed that the OFC model displays true criticality
only for a non-dissipative relaxation process. The apparent scale invariance observed in the 
avalanche size distributions produced from numerical simulations of the model is due to very long
(but finite) correlation length. True criticality is supposed to display an infinitely long 
correlation length. Thus, the issue of observing SOC in dissipative systems remains an open
question.

In this Letter, we derive a general SOBP formulation that is carried out by first 
generalizing the toppling rules: An active site ($z=1$)  may transfer two grains to two randomly
selected neighbors with probability $\alpha$ or, with probability $\beta$, may transfer only
one grain to a randomly chosen site while retaining the other, thus remaining active after
toppling. Most importantly, we show that criticality is preserved, even with the presence
of dissipation ($\alpha + \beta < 1$), in a dynamically evolving population characterized by
a birth rate $\eta$ of new active sites and mortality rate $\lambda$ of existing active sites, 
in contrast to previous claims ~\cite{lzs96}. The criticality of the sandpile is determined
by the demographic parameters rather than the $\alpha$ nor $\beta$.

\smallskip
\emph{Demographically dynamic SOBP}.--- 
The mean-field description of the sandpile model is defined by 
two stable heights, $z = 0$ (inactive or empty state) and $z=1$ (active or occupied state). 
An active site becomes unstable by the addition of
one grain, i.e., $z=2$. In a similar fashion as in~\cite{lzs95}, boundary conditions are 
implemented by allowing only $n$ generations of toppling. This allows the mapping of the 
sandpile model to a branching tree with $N= 2^{n+1}-1$ nodes. 

Let us define $p = N_1/N$ as the density of active ($z=1$) sites, where $N_1$ is the actual
number of active sites and $N$ is the total number of sites. This density is time-dependent
since $N_1$ changes in time, starting at $N_1 = 0$ without loss of generality. 
Each of the $(1-p)N$ inactive ($z=0$) sites has a 
probability $\lambda$ of being excited, whereas each of the $pN$ active sites has a
probability $\eta$ of being de-excited. With such process, the system can be described as
\emph{demographically dynamic}. This update rule initiates each time step.

Subsequently, an external energy grain is added to a 
randomly selected site. This grain has a probability $p$ of landing on an active site 
($z=1$) thereby rendering it unstable
($z=2$). When this occurs, the demographic process is momentarily frozen and relaxation
ensues. There are two ways that an unstable active site may relax: (1) with
probability $\alpha$ the site topples by transferring two grains to two randomly selected neighbors, and 
consequently becomes inactive; (2) with probability $\beta$ the site topples by transferring only one grain
to a randomly selected neighbor, and remains active after toppling. The topplings result to
an avalanche which propagates until all
the sites relax to stable states, after which the demographic process resumes.

Since neighbor selection is uniformly random, the avalanche can be approximated as a branching 
process~\cite{harris63} with branching probability 
$q_k = \alpha\delta_{k,2}+\beta\delta_{k,1}+(1-\alpha-\beta)\delta_{k,0}$. It follows that the 
generating function for the cluster size is
\begin{equation}
\label{eq:generating_function}
F(\omega) = \frac{1-b\omega-\sqrt{1-2b\omega+a\omega^2}}{2\alpha p}\,\, ,
\end{equation}
where $a=\beta^2p^2-4\alpha p\left(1-(\alpha+\beta)p\right)$ and $b=\beta p$. The branching
process is critical~\cite{harris63} if $\langle k\rangle = \sum kq_k = 1$ which corresponds to the condition
$p_c = (2\alpha+\beta)^{-1}$. It follows that this condition leads to $a=2b-1$.

The avalanche size distribution $P(s)$ is related to the generating function according to:
\begin{equation}
\label{eq:def_generating}
F(\omega) = \sum_{s=1}^{\infty} P(s)\omega^s\,\, .
\end{equation}
By expanding (\ref{eq:generating_function}) around $\omega=0$ and then comparing it with
(\ref{eq:def_generating}), we are able to derive a recurrence relation for the cluster
size distribution subject to the end conditions $P(0) = 0$ and $P(1) = (b^2-a)/4\alpha$:
\begin{equation}
\label{eq:recurrence_relation}
P(s) = \frac{b(2s-1)P(s-1)-a(s-2)P(s-2)}{s+1}\,\, .
\end{equation}

Graphically, Equation~(\ref{eq:recurrence_relation}) can be shown to define a power
law with exponent $\tau_s = 3/2$ for $s >> 1$ at the critical condition $a=2b-1$ with
$b\in [0,1)$. In particular, for $b=0$ the recurrence relation reduces to the analytic prediction
in~\cite{lzs96, pp03}, which only holds for odd values of $s$.

Dissipation in the model is 
implemented by allowing $q_{k=0}>0$ or equivalently, $\alpha + \beta < 1$. Hence, with probability 
$q_0=1-(\alpha+\beta)$, an unstable active site ($z=2$) dissipates two grains out of the system. 
In~\cite{lzs96} the dissipation probability $q_0 \equiv \epsilon$.
For conservative models, the amount of transferred grains is equal to the number of grains lost
by a relaxing site, at least on the average~\cite{vz97}, so that $\epsilon = 0$. 

Since we have embedded avalanche dynamics within a ``demographic" process, then we define
the relation $\lambda/\eta = 2\alpha+\beta-1$, 
which describes how avalanche dynamics is interspersed in a dynamic population. 
The demographic parameters $\eta$ and $\lambda$, and the branching parameters $\alpha$ 
and $\beta$ become interdependent. This condition can be equivalently written as,
\begin{equation}
\frac{\lambda}{\eta} = \frac{1-p_c}{p_c}\,\, ,
\label{eq:matching-condition}
\end{equation}
by recognizing that $p_c = (2\alpha+\beta)^{-1}$. 
Letting $\lambda\rightarrow 0$ and $\eta\rightarrow 0$ simultaneously is equivalent to the
limit $1-p_c\rightarrow p_c$ which implies that $p_c\rightarrow 1/2$. This is exactly the
result presented in~\cite{lzs95} which can only be achieved if $\alpha=1$ and $\beta=0$.
For this case, the stationary state is a SOC state only in the non-dissipative regime
($\alpha+\beta = 1$) as proven in~\cite{lzs96}.

\begin{figure}
\includegraphics[width=0.49\columnwidth]{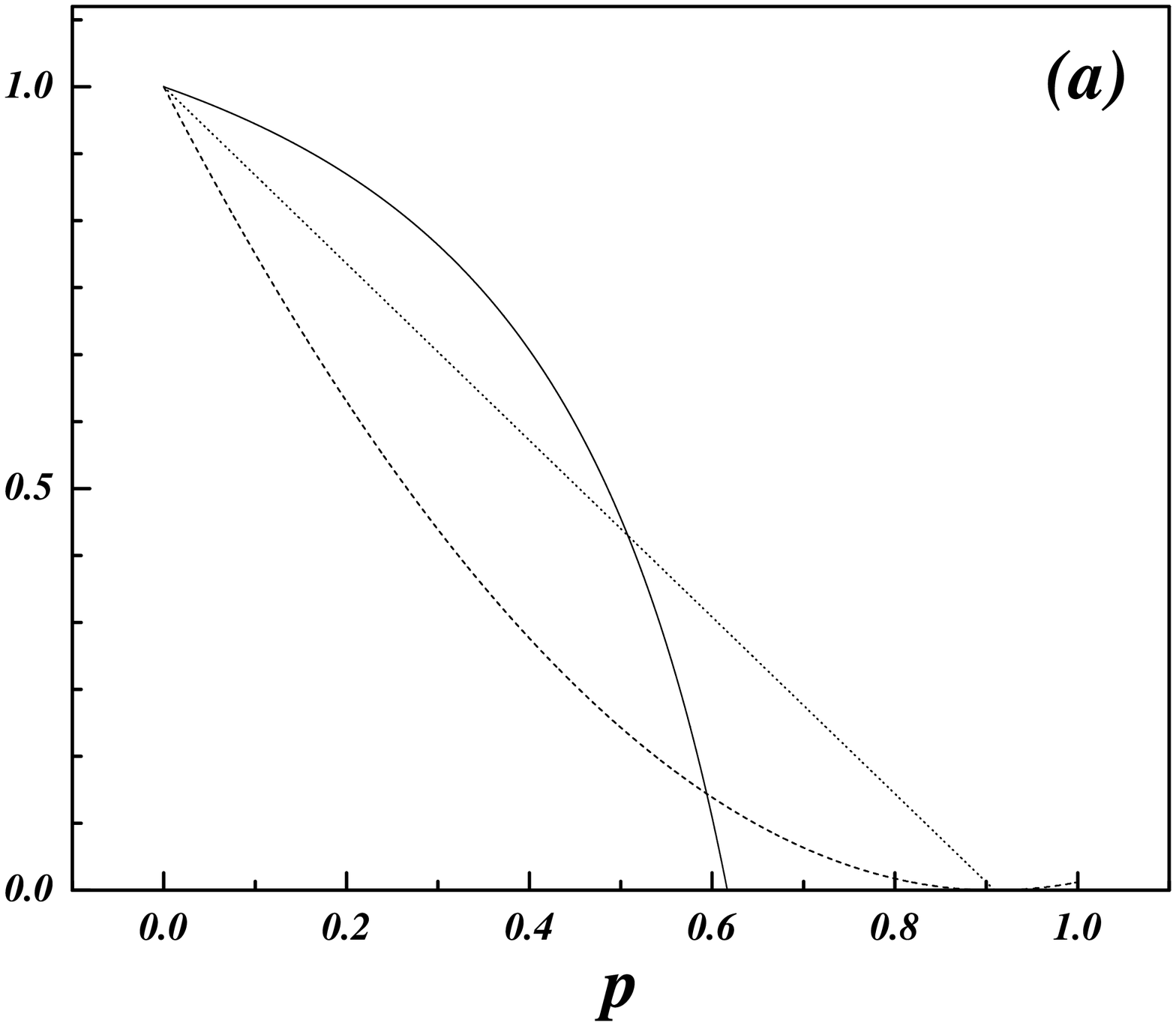}
\includegraphics[width=0.49\columnwidth]{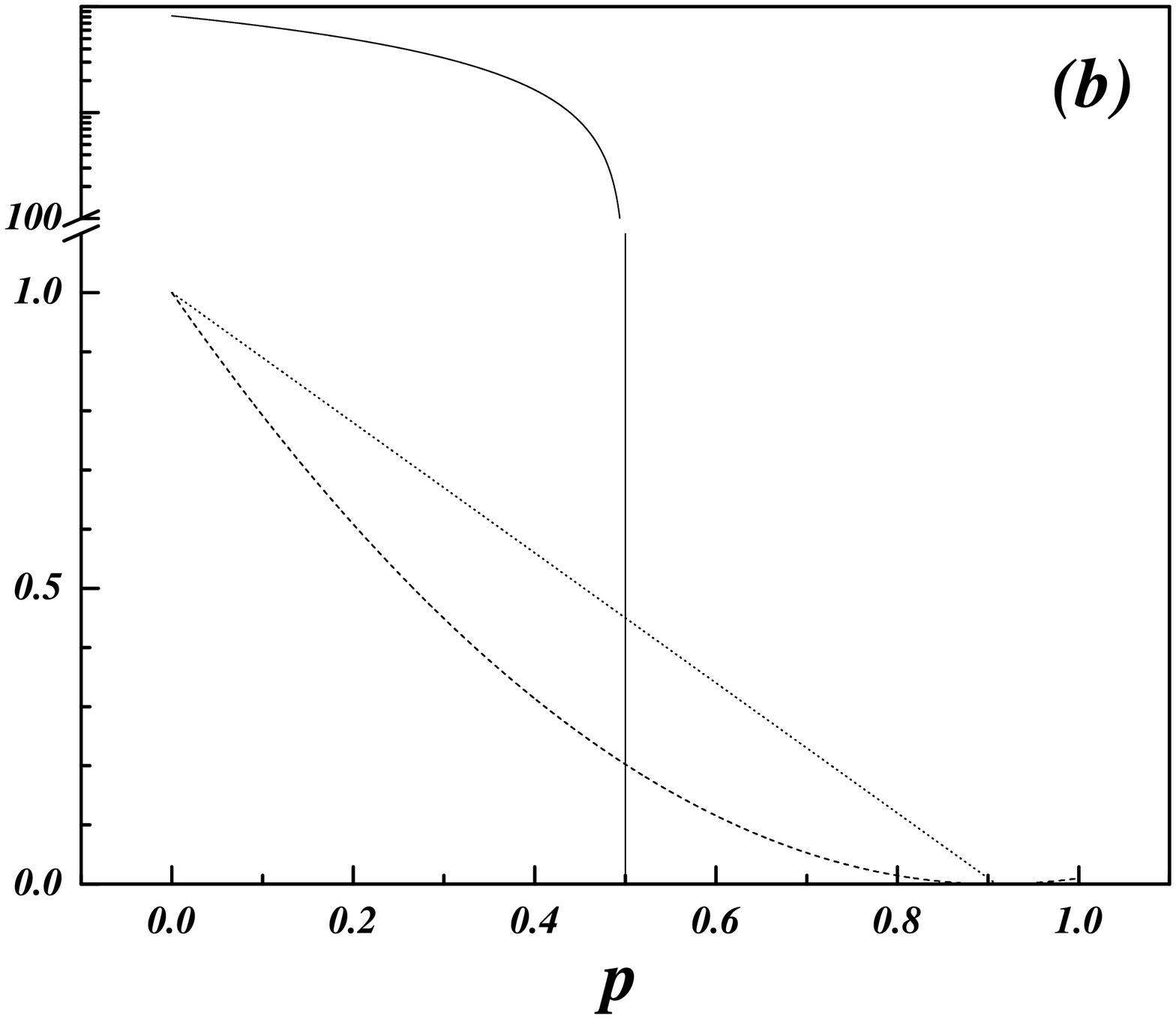}
\includegraphics[width=0.49\columnwidth]{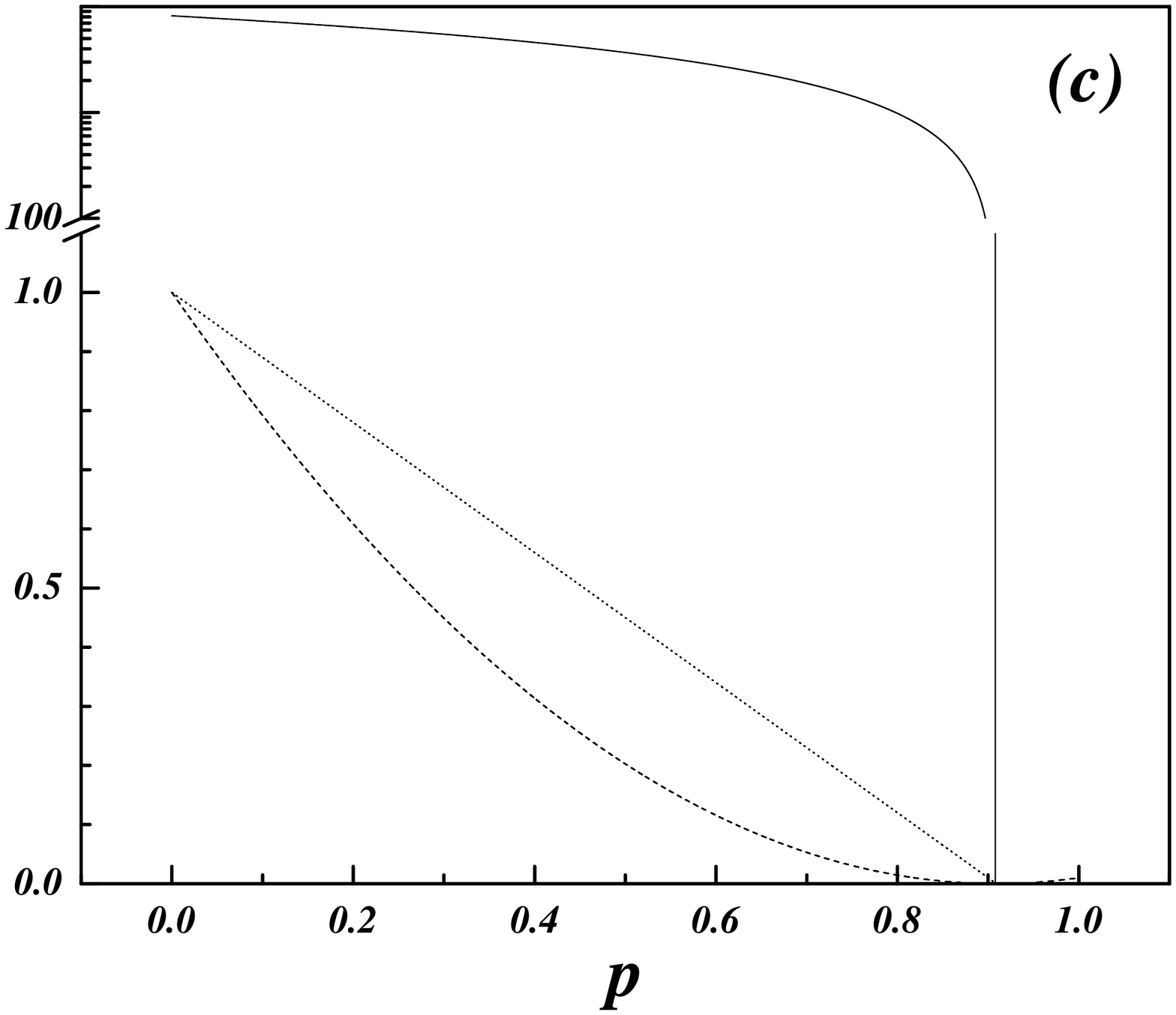}
\includegraphics[width=0.49\columnwidth]{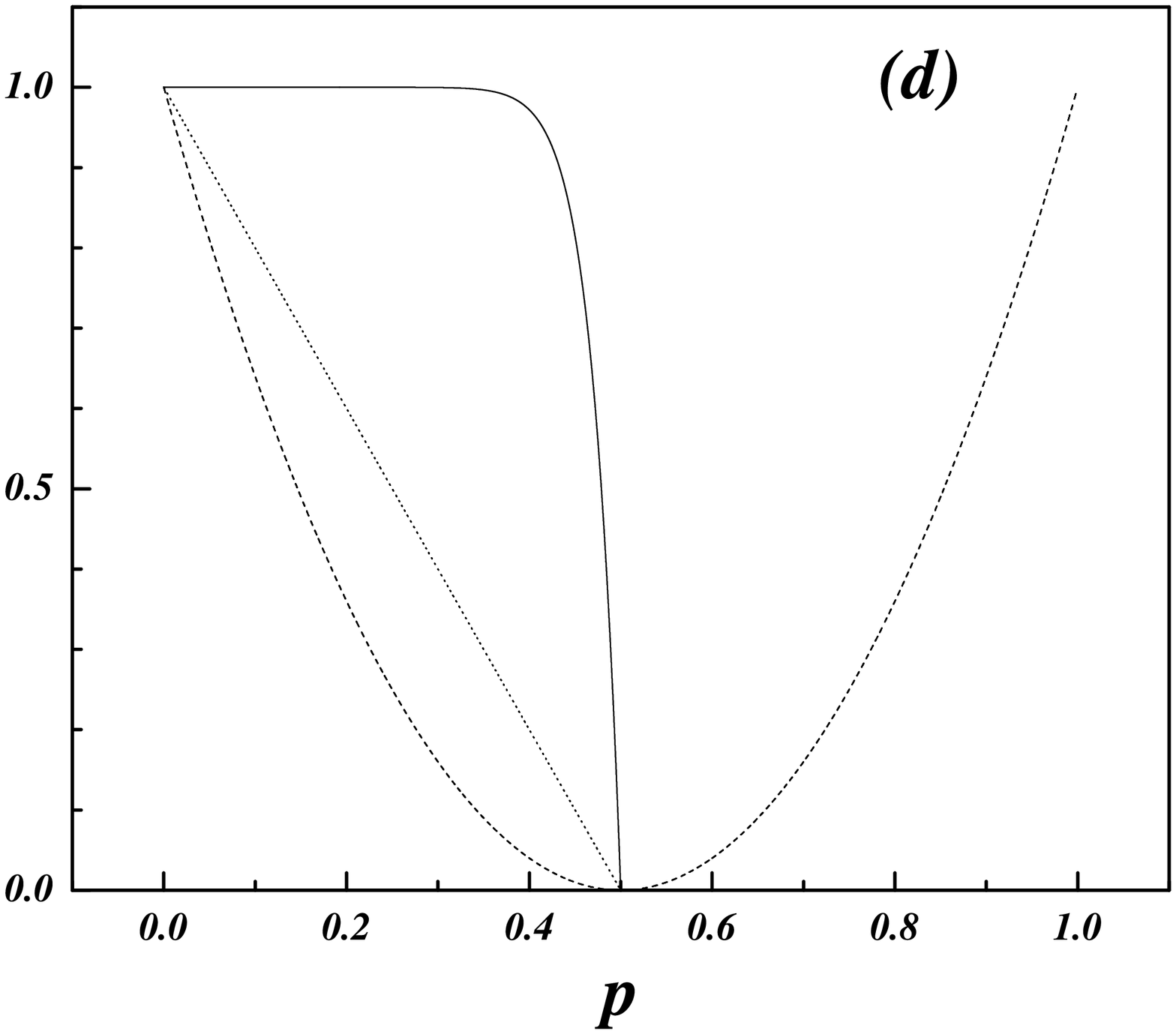}
\caption{
Rate of change $d p/d t$ of the density of active sites (solid lines),
criticality condition $1 + a - 2b$ (dashed lines) and the function
$1-(2\alpha+\beta)p$ (dotted lines) as a function of $p$. \textbf{(a)} 
 $\alpha=0.3$, $\beta=0.5$, $\eta=0$, (static population); \textbf{(b)}
 $\alpha=0.3$, $\beta=0.5$, $\eta=2^{-4}$, $\lambda=\eta$; \textbf{(c)} 
 $\alpha=0.3$, $\beta=0.5$, $\eta=2^{-4}$, $\lambda=(2\alpha+\beta-1)\eta$; \textbf{(d)}
 $\alpha=1.0$, $\beta=0.0$, $\eta=0$ (static population).
The self-organized branching process is critical if all three curves
mutually intersect at zero at $p=p_c$, which occurs for \textbf{(c)} and \textbf{(d)}.
}
\label{fig::criticality-conditions}
\end{figure}

\begin{figure} 
\includegraphics[width=1.0\columnwidth]{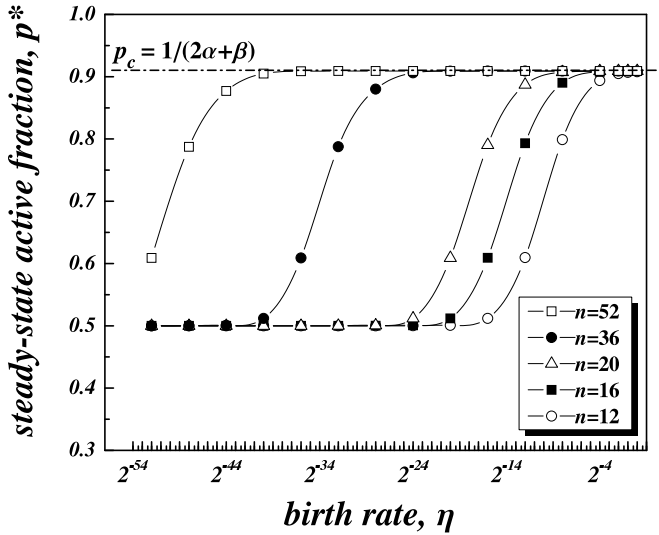}
\caption{
Steady state density of active sites $p^*$ as a function of excitation rate $\eta$
for different system dimension $n$.
Parameter values: $\alpha=0.55$, $\beta=0$.
The steady state is a subcritical state below a threshold excitation rate $\eta_{th}(n)$,
which decreases to zero with increasing system dimension $n$. Above $\eta_{th}$,
$p^* \rightarrow p_c$. A phase transition from a subcritical state to a critical 
state occurs at $\eta_{th}$. 
}
 \label{fig::phase-transition}
\end{figure}

\begin{figure}
\includegraphics[width=1.0\columnwidth]{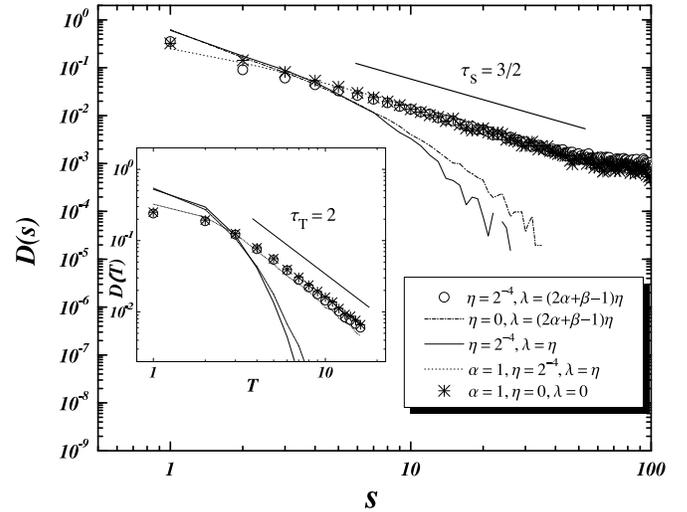}
\caption{
Numerical simulations of the demographically dynamic SOBP for system dimension $n=16$
after an integration time of $2^{17}$ iterations.
Avalanche size distribution for $\alpha=0.3$, $\beta=0.5$ (dissipative) and
$\alpha=1$, $\beta=0$ (non-dissipative). Self-organized 
criticality is achieved for a dissipative sandpile ($\alpha+\beta < 1$). At 
criticality, the distribution follows a power law with exponent $\tau_s=3/2$ that is
consistent with mean-field analysis. The inset graph shows the avalanche duration 
distribution that fits to a power law with exponent $\tau_T=2$ at criticality. 
}
\label{fig::avalanche-distributions}
\end{figure}

The rate of change of the density $p$ of active sites can be written as
\begin{widetext}
\begin{equation}
\label{eq:dp_dt}
\frac{d p}{d t} = (1-p)\eta  - p\lambda  + \frac{1}{N}\left(1-((2\alpha+\beta)p)^n  
		- \frac{(1-\alpha-\beta)p}{1-(\alpha+\beta)p}
		\left[1+\frac{1-((2\alpha+\beta)p)^{n+1}}{1-(2\alpha+\beta)p}
		-(2((2\alpha+\beta)p)^n)\right]\right)
		+ \frac{\sigma(p, t)}{N}
\end{equation} 
\end{widetext}

\noindent for which, $\sigma(p,t)/N$ is a noise term corresponding to the fluctuation around 
the stationary value of $p$ that vanishes with increasing system size $N$~\cite{lzs96}.
In order for the sandpile dynamics to be self-organized critical (SOC), the stationary state 
of the dynamical system must be a critical state. To show this, we solve for the fixed point
solution of Equation~(\ref{eq:dp_dt}).
Generally, the fixed-point solution of the density of active sites $p^*$ does not coincide with 
its critical value $p_c$. Moreover, $p^* < p_c$ implies that the steady state is a subritical state for which
the avalanche distribution possesses a characteristic scale and finite correlation length.
Shown in Fig.~\ref{fig::criticality-conditions} is a comparison of graph for $dp/dt$ and the critical conditions
$a-2b+1$ and $1-(2\alpha+\beta)p$ as functions of $p$. If the steady state is a critical
state, all three functions must intersect at zero at $p_c$. For a static population~\cite{lzs96},
$dp/dt = 0$ has a root $p^* < p_c$ unless $\epsilon = 1-\alpha-\beta = 0$. This implies that
self-organized criticality is only attained for a non-dissipative ($\epsilon=0$) system as
suggested in~\cite{lzs96}. However, in a dynamic population the critical state for which
$p_c = (2\alpha+\beta)^{-1}$ is reached above a threshold excitation rate $\eta_{th}$ that approaches to
zero as the system dimension $n\rightarrow\infty$. Illustrated in Fig.~\ref{fig::phase-transition}
is a transition from a subcritical steady-state to a critical steady-state for a degree of
dissipation $\epsilon = 0.2$ for which $\alpha=0.3$ and $\beta=0.5$.
With the coupling defined in Equation~(\ref{eq:matching-condition}), the mean number of new sites
gaining activity becomes equal to the mean number of active sites losing activity. Hence,
the critical state coincides with a stationary state, that is, $(1-p)\eta - p\lambda = 0$
in Equation~(\ref{eq:dp_dt}) after carrying out a substitution using Equation~(\ref{eq:matching-condition}). 
Conversely, for
a stationary state of the system wherein birth of activity equals death of activity, the 
SOBP can manifest criticality even in the presence of dissipation since $\alpha+\beta$ can
possibly be less than 1. The presence of a threshold excitation rate, $\eta_{th}$, is a requirement
of time-scale separation between the avalanche propagation and the population dynamics~\cite{vz97}.
We demonstrate our prediction with numerical simulations. The results are illustrated in
Figure~\ref{fig::avalanche-distributions}, clearly showing the avalanche size and duration
distributions with exponents $\tau_s = 3/2$ and $\tau_T = 2$ even with the presence of
dissipation.

Hence, a demographically dynamic population is capable of achieving the critical state
amidst the presence of local dissipation. This is significant for the fact that criticality has
been observed ubiquitously in systems which do not display any degree of local
conservation. Moreover, these systems possess population dynamics. 
Several attempts have been proposed to show nonconservative systems exhibiting self-organized
criticality. One intriguing example is the Olami-Feder-Christensen model~\cite{ofc92}, which
quite recently had been proven to display criticality only in the conservative regime~\cite{cp00}.

The idea
of coupling population dynamics with self-organized criticality is reminiscent of the forest-fire
model. But unlike the forest-fire model, the driving rates $\eta$ and $\lambda$
do not have to satisfy the limits $\lambda\rightarrow 0$, $\eta\rightarrow 0$, and 
$\lambda/\eta\rightarrow 0$ in order to
display criticality in the system. As long as these demographic parameters satisfy a matching condition
(i.e., $\lambda/\eta = (1-p_c)/p_c$, wherein $p_c = (2\alpha + \beta)^{-1}$, criticality is possible
to achieve for $\eta > \eta_{th}$.

\begin{figure}
\includegraphics[width=1.0\columnwidth]{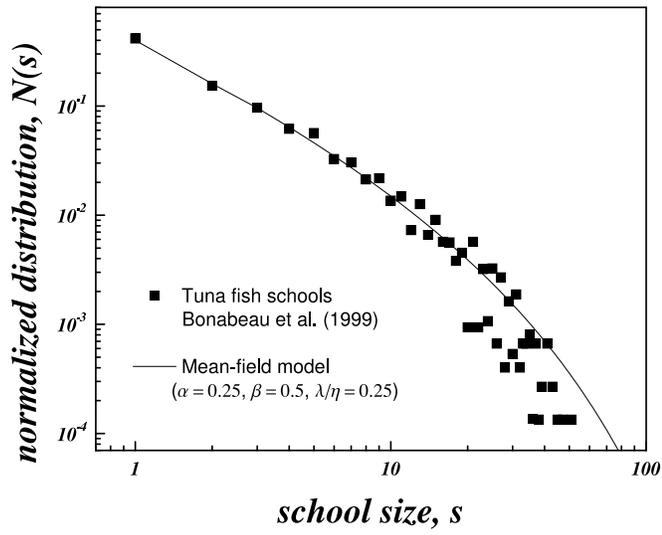}
\caption{
Normalized fish school size distribution data from~\cite{bonabeau99} fitted with 
Eq.~(\ref{eq:recurrence_relation}): $\alpha=0.25$, $\beta=0.5$, $\lambda/\eta=0.25$.
The truncation at the upper tail of the distribution is evidence of subcriticality
owing to significant probability of deexcitation ($\lambda=0.25\eta$).
}
\label{fig:bonabeau-data}
\end{figure}

\smallskip
\emph{Fish school size scaling}.--- Bonabeau et al.~\cite{bonabeau99} presented data of aggregation
of fishes and formulated an elementary model of animal aggregation. In their mean-field model, they
considered an interplay between aggregation and splitting. These parameters are analogous to the 
excitation rate $\eta$ and death rate $\lambda$. In Figure~\ref{fig:bonabeau-data}, the data for tuna fish schools is
fitted with the mean field solution (Equation~(\ref{eq:recurrence_relation})). 
The school size distribution manifests subcriticality owing to
the mismatch between the demographic parameters and the branching parameters. 

Adding a grain to a sandpile is analogous to stimulus reception of an ensemble of fishes. When a fish
(or a small group of fishes) from the ensemble is excited by the stimulus it passes the information
to other fishes. An information cascade ensues eventually synchronizing the action of the fishes
via a proposed clustering biomechanism known as \emph{allelomimesis}~\cite{juanico03} from which emerges
aggregation to a food source as an example. Allelomimesis has been found to be an appropriate mechanism
that produces scaling in a wide sample of animal and human group size distributions~\cite{juanico05}.
The subsequent stimulation of other fishes due to the 
observance of prior stimulated fishes can be thought of as a branching process wherein a stimulated
fish has an average probability $q_k$ of stimulating $k$ other fishes (at random). With a probability
$q_{k=0}>0$, the relaxation is dissipative~\cite{lzs96}. Unstimulated fishes ($z=0$) of the 
ensemble may spontaneously get stimulated at an average rate of $\eta$. On the other hand, 
stimulated fishes ($z=1$) may spontaneously get deexcited at
an average rate of $\lambda$. The dissipative nature of fish clustering is due to an imperfect tendency
to copy neighbors as shown by Juanico et al.~\cite{juanico03,juanico05} through an agent-based model
of clustering by allelomimesis.

The resulting cluster size distribution was found to behave as a power-law with an exponent $3/2$.
However, actual data exhibits a characteristic scale in the distribution (subcriticality) that manifest as a 
cutoff size~\cite{bonabeau99}. In Bonabeau et al.'s model, the truncation of the distribution is
attributed to a splitting parameter $\sigma$ which constrains the lifetime of the fish school by
virtue of breaking up after a certain amount of time. This splitting parameter is analogous to 
deexcitation in our model that happens with an average probability $\lambda$. Subcriticality that
manifest in the cluster size distribution is interpreted as the condition $\lambda/\eta > (1-p_c)/p_c$.

The main drawback with present SOC models is the need for some type of local conservation as an 
essential ingredient for a system to display SOC. The existence of SOC in nonconservative models
would be highly desirable in this context, since, in practice, some type of dissipation is always 
present in nature~\cite{cp00}.

\bibliography{prl-juanico-2006}
\end{document}